\documentclass[aps,showpacs,onecolumn]{revtex4}
\usepackage{amssymb}
\usepackage{graphicx}
\usepackage{subfigure}
\usepackage[dvipdfm]{hyperref}

\begin{document}

\title{Numerical Studies of Quantum Hall Ferromagnetism in Two-Subband
Systems}
\author{Xiao-Jie Hao$^{(1)}$}
\author{Tao Tu$^{(1)}$}
\email{tutao@ustc.edu.cn}
\author{Yong-Jie Zhao$^{(1)}$}
\author{Guang-Can Guo$^{(1)}$}
\author{H. W. Jiang$^{(2)}$}
\author{Guo-Ping Guo$^{(1)}$}
\email{gpguo@ustc.edu.cn}
\affiliation{$^{(1)}$ Key Laboratory of Quantum Information, University of Science and
Technology of China, Chinese Academy of Sciences, Hefei 230026, People's
Republic of China \\
$^{(2)}$ Department of Physics and Astronomy, University of California at
Los Angeles, 405 Hilgard Avenue, Los Angeles, CA 90095, USA}

\begin{abstract}
We carry out a numerical study of the quantum Hall ferromagnetism in  a
two-subband system using a set of experimental parameters in a  recently
experiment [X. C. Zhang, I. Martin, and H. W. Jiang, Phys.  Rev. B \textbf{74%
}, 073301 (2006)]. Employing the self-consistence  local density
approximation for growth direction wave function and  the Hartree-Fock
theory for the pseudospin anisotropy energy, we are  able to account for the
easy-axis and easy-plane quantum Hall  ferromagnetism observed at total
filling factor $\nu = 3$ and $\nu=  4$, respectively. Our study provides
some insight of how the  anisotropy energy, which highly depends upon the
distribution of  growth direction wave functions, determines the symmetry of
the  quantum Hall ferromagnets.
\end{abstract}

\date{\today}
\maketitle

\baselineskip20pt

\section{Introduction}

\label{sec:introdunction} Multi-component quantum Hall systems have
exhibited a collection of interesting phenomena which are the manifestation
of electronic correlations \cite{review}. Such correlations become
particularly prominent when two or more sets of Landau levels (LLs) with
different layer, subband, valley, spin, or Landau level indices are brought
into degeneracy \cite{Eisenstein1994,KunYang1994,Wescheider,Shayegan}. In
experimental systems, different LLs can be tuned to cross by varying gate
voltage, charge density, magnetic field or the magnetic field tilted angle
to the sample. One of the attractions is the formation of quantum Hall
ferromagnets (QHFs) due to the exchange interactions of the two subbands
states, termed as pseudospins \cite{QHF,MacDonald1990}. Self-consistent
local density approximation (SCLDA) and Hartree-Fock mean field method can
be performed on the calculation of ground state energy and quasi-particle
energy gap \cite{Wescheider,MacDonald2000,Shayegan1998,MacDonald1990}.

Recent experiments in single quantum well with two-subband occupied systems
\cite{Hirayama,Jiang2006}, showed evidence of QHFs when two LLs were brought
into degeneracy. The QHFs can either be easy-axis or easy-plane, depending
on the details of LL crossing configurations \cite{Hirayama,Jiang2006}. In
this paper we follow the theoretical framework of Jungwirth and MacDonald
\cite{MacDonald2000} and numerically calculate the pseudospin anisotropy
energy using the sample parameters in the experiment of Zhang \textit{et al.}
\cite{Jiang2006}. The result confirms the QHFs taking place at total filling
factor $\nu =3$ and $\nu =4$ are expected to be easy-plane and easy-axis
QHFs, respectively. As we will discuss in this paper, the gate bias voltage,
which affects the spatial distributions of the wave functions of the two
subbands, plays a leading role in the formation of easy-plane and easy-axis
QHFs.

\section{Pseudospin Quantum Hall Ferromagnets}

\label{sec:pseud-quant-hall}

\begin{figure}[ht]
\includegraphics[width=0.7\columnwidth]{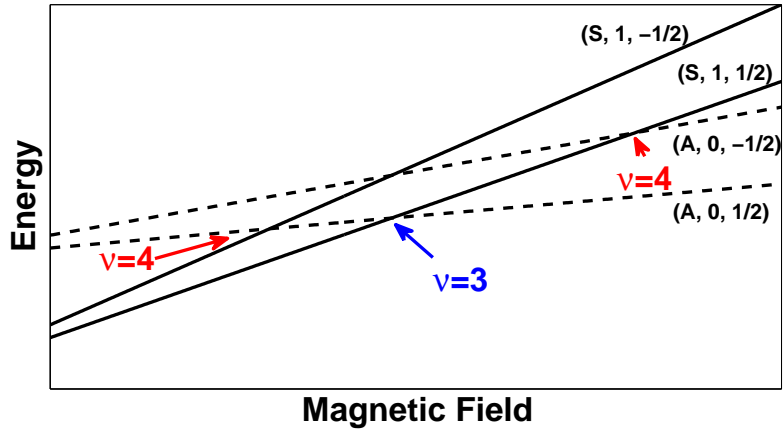}
\caption{Landau level diagram of two-subband system. Four Landau levels
labeled by $(\protect\xi, n, s)$ are crossing to each other during the
increasing of the magnetic field (see text for details). Three arrows point
out the crossing point of Landau levels at filling factor $\protect\nu=3$
and $\protect\nu=4$. }
\label{fig:Landau}
\end{figure}

The pseudospin representation is used to describe the valence LLs
degenerated in the Fermi level \cite{MacDonald1990,MacDonald2000}. Following
the theoretic studies of pseudospin QHFs \cite{MacDonald1990,MacDonald2000},
here we focus on a two-subband two dimensional electron system, in which the
LLs are labeled by $(\xi ,n,s)$, where $\xi =S/A$ is first/second subband
(in the no biased quantum well, also called symmetry/antisymmetry subband)
index, $n=0,1,\cdots $ is Landau level in-plane orbit radius quantum number,
and $s=\pm \frac{1}{2}$ represents real spin. When two LLs are brought close
to degeneracy but still sufficiently far from other LLs, one of them can be
labeled as pseudospin up ($\sigma =\Uparrow $) and the other as pseudospin
down ($\sigma =\Downarrow $). In the experimental work of Zhang \textit{et
al.} \cite{Jiang2006}, as shown in Fig.~\ref{fig:Landau}, we can label
pseudospin up ($\sigma =\Uparrow $) as $(S,1,\frac{1}{2})$ and pseudospin
down ($\sigma =\Downarrow $) as $(A,0,\frac{1}{2})$ at filling factor $\nu =3
$. At filling factor $\nu =4$, we label pseudospin up ($\sigma =\Uparrow $)
as $(S,1,\mp \frac{1}{2})$ and pseudospin down ($\sigma =\Downarrow $) as $%
(A,0,\pm \frac{1}{2})$. Here the upper and lower signs before spin index
refer to the crossing point at lower and higher magnetic field, respectively.

At this moment, we would emphasize the essential similarity between
two-subband system and ordinary bilayer system (double quantum well). On one
hand, due to a separation of charges to opposite sides of the well
originating from the Coulomb repulsion, a wide single quantum well can be
modeled as an effective bilayer system \cite%
{Wescheider,MacDonald2000,Abolfath}. Therefore in a single quantum well
two-subband structure, the states of electrons can be characterized by two
parameters similar to a bilayer system: the tunneling gap $\Delta _{SAS}$
and effective layer separation $d$. The $\Delta _{SAS}$ is chosen as equal
to the difference between the lowest two subband energy levels and $d$ is
given by the distance between two centers of the charge distribution in this
single quantum well. On the other hand, let us assume the two dimensional
electron gas is in the $x-y$ plane in the following calculation and
discussion, so the sample growth direction is aligned with the $z$ axis. In
a no biased bilayer system, although the layer wave function in $z$
direction is spatially separated in two layers, once included in the effect
of tunneling between two layers $\Delta _{t}$ and rediagonalized the $z$
direction Hamiltonian, it will be just like the two-subband system which has
symmetry and antisymmetry subbands. And the symmetry-antisymmetry gap $%
\Delta _{SAS}$ is equal to the inter-layer tunneling $\Delta _{t}$. Even
when a bias voltage between two layers exists, the circumstance is more or
less the same. Since they are theoretically identical, all the pseudospin
language theory used in bilayer system is available in the two-subband
system.

While pseudospin up and pseudospin down LLs are degenerate but the number of
electrons is not enough to fill all the two LLs, electrons will stay in a
broken-symmetry ground state. Actually, the state electrons choosing is a
linear combination of two pseudospin LLs which minimizes the system total
energy. Typically, the many-body ground state can be written as follows \cite%
{MacDonald2000}:
\begin{equation}  \label{eq:many-body-groud}
\left\vert \Psi\lbrack \hat{m}\rbrack \right\rangle =\prod_{k=1}^{N_{\phi
}}c_{\hat{m},k}^{\dagger}|0\rangle,
\end{equation}
where $c_{\hat{m},k}^{\dagger }$ creates the single-particle state oriented
in a certain unit vector $\hat{m}=(\sin{\theta}\cos{\varphi}, \sin{\theta}%
\sin{\varphi},\cos{\theta})$ with wave function:
\begin{equation}  \label{eq:groud-state}
\psi _{\hat{m},k}(\vec{r})=\cos \big(\frac{\theta }{2}\big)\psi _{\Uparrow
,k}(\vec{r})+\sin \big(\frac{\theta }{2}\big)e^{i\varphi }\psi _{\Downarrow
,k}(\vec{r}).
\end{equation}
Here $\psi _{\sigma ,k}(\vec{r})$ is the single-particle state wave function
which contains growth direction subband wave function $\lambda_{\xi}(z)$ and
in-plane LL wave function $L_{n, s, k}(x, y)$:
\begin{equation}
\psi _{\sigma ,k}(\vec{r}) =\lambda _{\xi }(z)L_{n,s,k}(x, y).
\label{eq:single-particle}
\end{equation}

Then the Hartree-Fock energy of the system can be obtained as following \cite%
{MacDonald2000}:
\begin{equation}
E_{HF}(\hat{m})\equiv \frac{\langle \Psi \lbrack \hat{m}]|H|\Psi \lbrack
\hat{m}]\rangle }{N_{\phi }}=-\sum_{i=x,y,z}\left( E_{i}-\frac{1}{2}U_{%
\mathbf{1},i}-\frac{1}{2}U_{i,\mathbf{1}}\right) m_{i}+\frac{1}{2}%
\sum_{i,j=x,y,z}U_{i,j}m_{i}m_{j}.  \label{eq:HF_energy}
\end{equation}%
Whether the ground state is easy-axis or easy-plane only depends on the
quadratic coefficient in pseudospin magnetization $m_{i}$ (i.e. the
pseudospin anisotropy energy $U_{xx}$, $U_{yy}$ and $U_{zz}$ ):
\begin{equation}
U_{ij}=\frac{1}{4}\int \frac{d^{2}\vec{q}}{(2\pi )^{2}}v_{ij}(0)-\frac{1}{4}%
\int \frac{d^{2}\vec{q}}{(2\pi )^{2}}v_{ij}(\vec{q}).
\label{eq:anisotropy_energy}
\end{equation}%
The first $\vec{q}=0$ term in Eq.~(\ref{eq:anisotropy_energy}) is the
Hartree term and the second term is the Fock term. $v_{ij}(\vec{q})$ can be
expanded to sum of several pseudospin matrix elements $v_{\sigma
_{1}^{\prime },\sigma _{2}^{\prime },\sigma _{1},\sigma _{2}}(\vec{q})$,
which are products of subband and the in-plane parts \cite{MacDonald2000}.
If $U_{zz}<U_{xx}=U_{yy}$, the system is in easy-axis QHF, which means the
pseudospin magnetization $\hat{m}$ is aligned either up or down; and if $%
U_{zz}>U_{xx}=U_{yy}$, the system is in easy-plane QHF, which is a coherent
superposition of the two pseudospin LLs.

\section{Growth Direction Wave Function Calculation Using SCLDA}

\label{sec:lda} We numerically calculate the growth direction ($z$
direction) wave function $\lambda_{\xi}(z)$ in Eq.~(\ref{eq:single-particle}%
) using SCLDA to compare the pseudospin anisotropy energy $U_{xx}$ and $%
U_{zz}$ in Eq.~(\ref{eq:HF_energy}) \cite{LDA}. In the SCLDA method, wave
function $\lambda(z)$ is described by the Schr\"{o}dinger equation:
\begin{equation}  \label{eq:schrodinger}
\left(-\frac{1}{2m^{*}} \frac{\partial^{2}}{\partial z^{2}} + V_{b}(z) +
V_{gate}(z) + V_{xc}(z) + V_{H}(z) \right) \lambda_{i}(z) =
\varepsilon_{i}\lambda_{i}(z).
\end{equation}
Here $m^{*}$ is the effective electron mass in GaAs, $V_{b}$ corresponds to
the conduction band discontinuity, $V_{gate}$ is the bias potential caused
by the difference of front and back gate voltage $\left\vert\Delta
V_{g}\right\vert$, $V_{xc}$ refers to the exchange-correlation potential
related to the electron charge distribution $n(z)$ (We use the form of $%
V_{xc}$ given by Hedin and Lundqvist \cite{ExchangeCorrelation}). The
Hartree term $V_{H}$ due to electrostatic potential is given in the Poisson
equation:
\begin{equation}  \label{eq:poisson}
V_{H}(z)=-\frac{2\pi e^{2}}{\epsilon} \int dz^{\prime}\left\vert
z-z^{\prime}\right\vert n(z^{\prime}).
\end{equation}

The subband energies $\varepsilon_{\xi}$, wave functions $\lambda_{\xi}(z)$
and the electron charge distributions $n_{\xi}(z)$ of both subbands can be
calculated by solving the Schr\"{o}dinger equation Eq.~(\ref{eq:schrodinger}%
) and the Poisson equation Eq.~(\ref{eq:poisson}) simultaneously \cite%
{Algorithm}.

\section{Numerical Result and Discussion}

\label{sec:numer-result-dissc} Taking the unit of energy as $e^{2}/\epsilon
l_{B}$ ($l_{B}$ is the magnetic length), we calculate the pseudospin
anisotropy energy $U_{xx}$, $U_{yy}$ and $U_{zz}$ at filling factor $\nu=3$
and $\nu=4$ in Zhang \textit{et al.}'s work \cite{Jiang2006}. At total
filling factor $\nu=3$, $U_{xx} \equiv U_{yy}=-0.370<U_{zz}=0.017$, so the
system will stay in easy-plane QHF. At one degenerate point of total filling
factor $\nu=4$, $U_{xx} \equiv U_{yy}=0>U_{zz}=-0.1266$, and at the other
degenerate point $U_{xx} \equiv U_{yy}=0>U_{zz}=-0.1293$. Both of the energy
differences of $U_{zz}-U_{xx}$ at $\nu=4$ indicate easy-axis QHF ground
states.

\begin{figure}[htb]
\subfigure [$\nu=3$]  {\label{fig:v3_phase}\includegraphics[width=0.45%
\columnwidth]{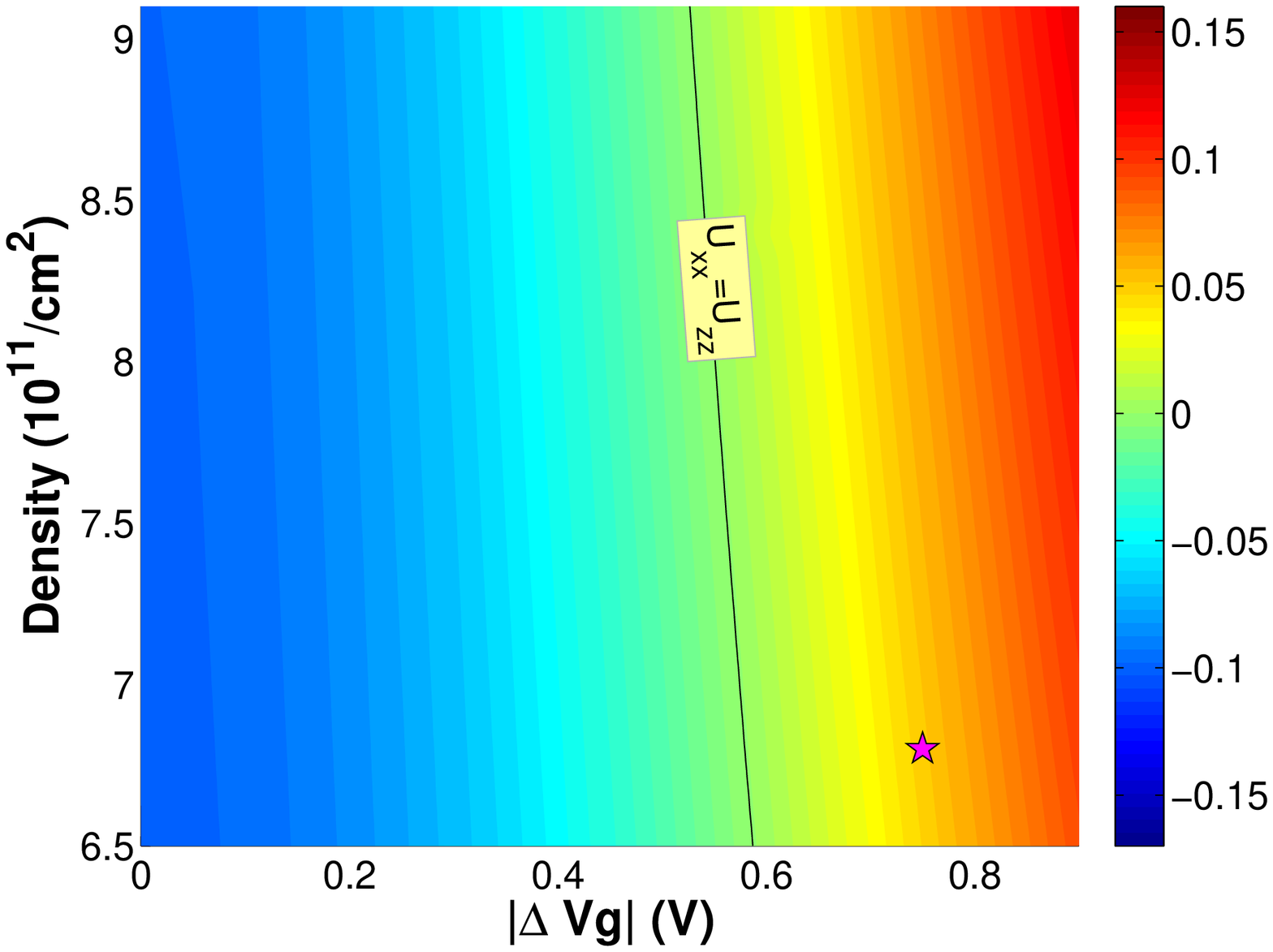}}  \subfigure [$\nu=4$]  {\label{fig:v4_phase}%
\includegraphics[width=0.45\columnwidth]{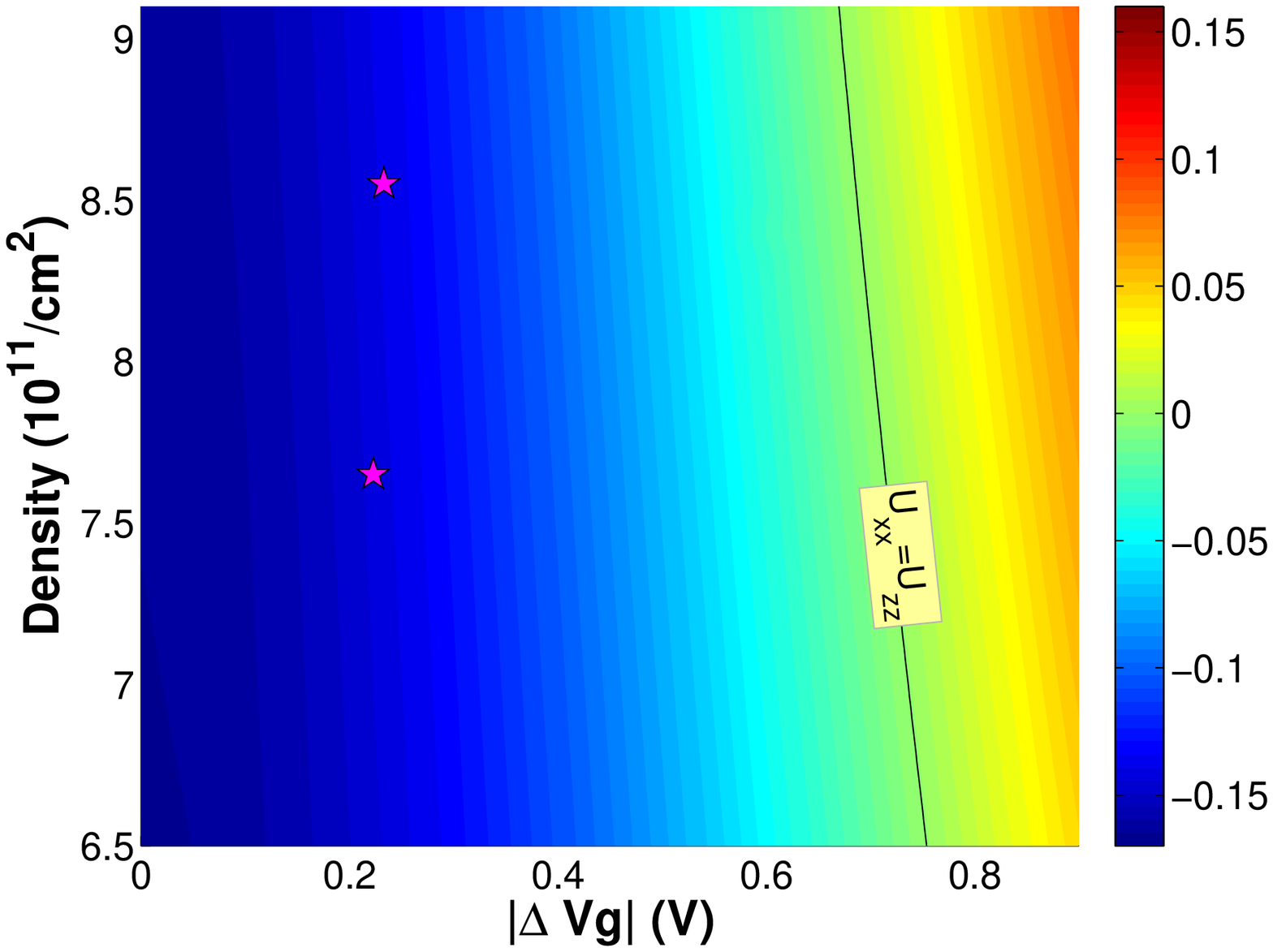}}
\caption{(Color online) Phase diagram of $U_{zz}-U_{xx}$ (unit: $e^{2}/%
\protect\epsilon l_{0}$)at filling factor (a) $\protect\nu=3$ and (b) $%
\protect\nu=4$. In the $U_{zz}-U_{xx}<0$ (left/blue) region easy-axis QHF is
energetically favorable while in the $U_{zz}-U_{xx}>0$ (right/red) region,
easy-plane QHF is favorable. Black lines in each figure labeled by $%
U_{xx}=U_{zz}$ show the critical positions where a quantum phase transition
from easy-axis to easy-plane QHF occurs. The three stars in these two
figures denote the $\protect\nu=3$ and $\protect\nu=4$ experimental
parameter in the experimental work of Zhang \textit{et al.}. }
\label{fig:phase}
\end{figure}

To illustrate the evolution from easy-axis QHF to easy-plane QHF, we
calculate the phase diagrams of $U_{zz}-U_{xx}$ as a function of bias gate
voltage $\left\vert\Delta Vg\right\vert$ and total density $n$ at filling
factor $\nu=3$ (Fig.~\ref{fig:v3_phase}) and $\nu=4$ (Fig.~\ref{fig:v4_phase}%
). In the following discussion, in order to make consistency, we choose $%
e^{2}/\epsilon l_{0}$ ($l_{0}=10 nm$) as the unit of energy. In the phase
diagram (Fig.~\ref{fig:phase}), we label the density and bias voltage $%
\left\vert\Delta Vg\right\vert$ position where the crossing occurs in the
experimental work of Zhang \textit{et al.} \cite{Jiang2006}. In the
left/blue (right/red) parts of each figure (Fig.~\ref{fig:v3_phase} and Fig.~%
\ref{fig:v4_phase}), where $U_{zz}< (>) U_{xx}$, easy-axis (easy-plane) has
the lower electron interaction energy. From Fig.~\ref{fig:phase}, we find
that the anisotropy energy difference $U_{zz}-U_{xx}$ is very sensitive to
the bias voltage $\left\vert\Delta Vg\right\vert$. If we could vary the gate
voltage across the black line labeled $U_{xx}=U_{zz}$ from left to right in
a determined density, a quantum phase transition from easy-axis to
easy-plane QHF will happen.

\begin{figure}[ht]
\includegraphics[width=0.7\columnwidth]{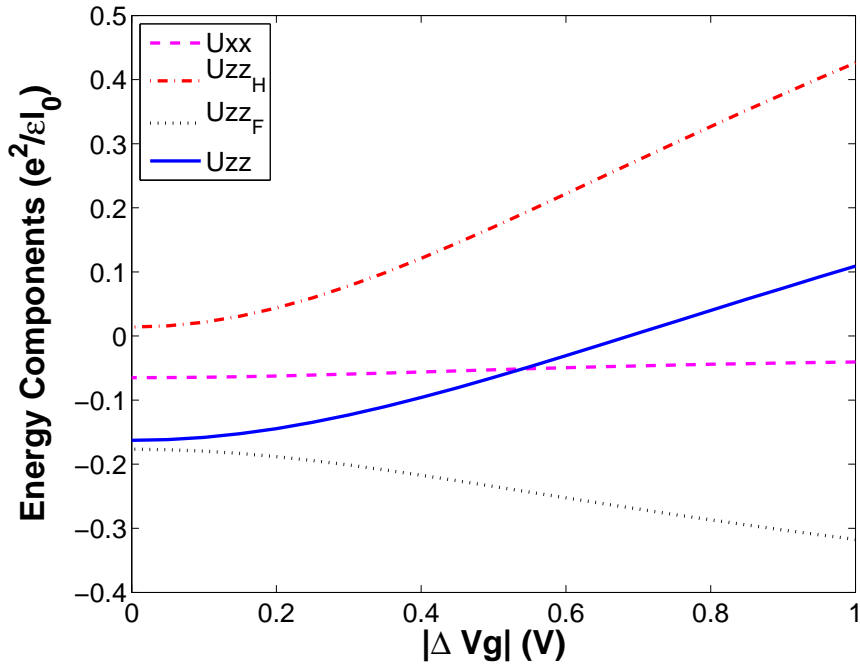}
\caption{Energy components of anisotropy energy $U_{xx}$ and $U_{zz}$ v.s.
bias voltage $\vert\Delta Vg\vert$. Except for $U_{xx}\equiv 0$ at $\protect%
\nu=4$, all the other terms are same at filling factor $\protect\nu=3$ and $%
\protect\nu=4$. $Uzz_{H}$ and $Uzz_{F}$ correspond to the Hartree term and
Fock term in the Hartree-Fock calculation of $U_{zz}$. }
\label{fig:energy}
\end{figure}

In order to make a comparison of the effect of each term in anisotropy
energy, we plot them in Eq.~(\ref{eq:anisotropy_energy}) as a function of $%
\left\vert\Delta Vg\right\vert$ at a certain density of $8.5%
\times10^{11}/cm^{2}$ in Fig.~\ref{fig:energy}. Here we have to point out
that all the formulas at $\nu=3$ and $4$ are the same, except that $U_{xx}$
vanishes at $\nu=4$. So the Fig.~\ref{fig:energy} is applicable to both
filling factor $\nu=3$ and $\nu=4$ only if in the $\nu=4$ situation the $%
U_{xx}$ term is set to a constant zero. The reason for the different
behavior of $U_{xx}$ at filling factor $\nu=3$ and $\nu=4$ will be discussed
in the following. Note that the dominated term in $U_{zz}-U_{xx}$ is the
Hartree term $Uzz_{H}$, which is due to the electrostatic potential of
electrons. It shows that the $Uzz_{H}$ term increases immediately with the
increasing bias voltage $\left\vert\Delta Vg\right\vert$.

\begin{figure}[htb]
\includegraphics[width=0.7\columnwidth]{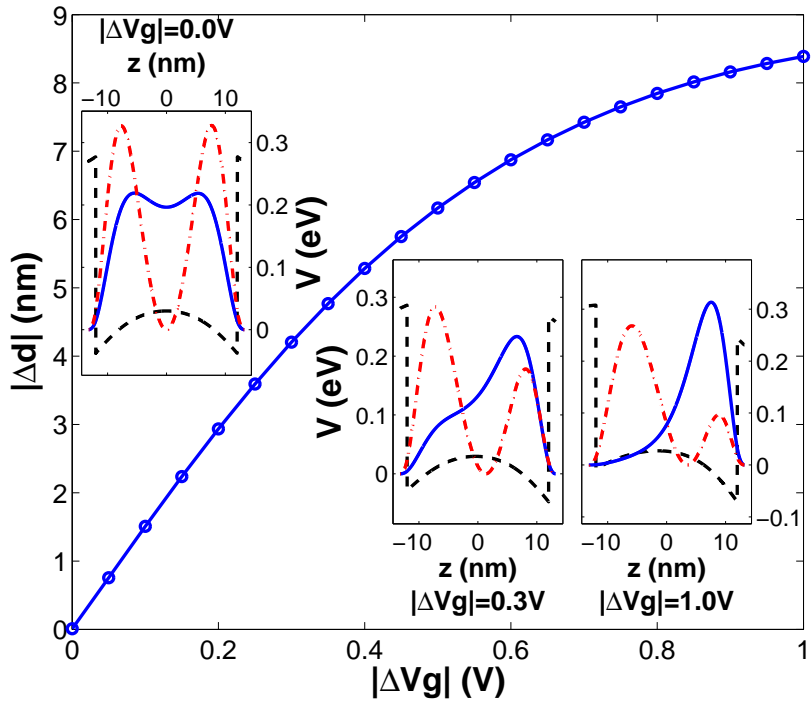}
\caption{(Color online) Effective subband separation $\left\vert\Delta
d\right\vert$ as a function of bias gate voltage $\left\vert\Delta
V_{g}\right\vert$ at the density of $8.5\times10^{11}/cm^{2}$. Three insets
show the well profiles (black dashed lines) with first (blue solid lines)
and second (red dash-dot lines) subbands wave functions at different bias
voltage: $\left\vert\Delta V_{g}\right\vert = 0.0 V$, $0.3 V$, and $1.0 V$.}
\label{fig:d_Vg}
\end{figure}

To examine what the role the bias voltage plays in the determination of
easy-plane or easy-axis QHF, we give some SCLDA results in the Fig.~\ref%
{fig:d_Vg}. We define an effective subband separation $\Delta d$ to describe
the distance between the cores of first and second subbands wave functions.
The $\left\vert \Delta d\right\vert $ as a function of bias voltage is
plotted in Fig.~\ref{fig:d_Vg}, also with some demonstrations of quantum
well configurations and subbands wave functions in different bias voltages ($%
\left\vert \Delta V_{g}\right\vert =0.0V$, $0.3V$, and $1.0V$). The data in
Fig.~\ref{fig:d_Vg} are all selected from the same density $8.5\times
10^{11}/cm^{2}$ as in Fig.~\ref{fig:energy}. It is obvious that the bias
gate voltage changes the effective separation of two subbands while changing
the well potential $V_{gate}(z)$ in Eq.~(\ref{eq:schrodinger}). The larger
the bias voltage is added, the farther the lowest two subbands are
separated. Since at a well separated $z$ direction wave function
configuration, all the electrons near the Fermi level filling in the same
subband, i.e. the same pseudospin level, will raise a larger electrostatic
energy, the easy-axis QHF is not favorable. Thus an easy-plane QHF can save
more Hartree energy in the larger bias voltage situation. On the other hand,
when the potential of the quantum well maintains a good symmetry in the
small bias voltage limit, electrons can pick up one of the two pseudospin
levels to keep the Hartree energy minimal and avoid the energetic penalty
from inter-subband tunneling as well.

As mentioned before, the difference of $U_{xx}$ between total filling factor
$\nu=3$ and $\nu=4$ could be explained in the similar way. The anisotropy
energy $U_{xx}$ or $U_{yy}$ is constituted of Hartree part and Fock part
(see Eq.~(\ref{eq:anisotropy_energy})). In our numerical calculation, we
find that the Hartree terms of $U_{xx}$ and $U_{yy}$ are always zero at both
$\nu=3$ and $\nu=4$. And the only non-zero term in $U_{xx}$ or $U_{yy}$ is
the Fock term $Uxx_{F} \equiv Uyy_{F}<0$ at filling factor $\nu=3$, which
owes to the exchange interaction. It implies that at filling factor $\nu=4$,
where pseudospin up $\sigma=\Uparrow(S, 1, \mp\frac{1}{2})$ and pseudospin
down $\sigma=\Downarrow(A, 0, \pm\frac{1}{2})$ have opposite real spins, the
easy-plane anisotropy, in which the electrons stay in both subband equally,
would cost much more exchange energy. But at $\nu=3$ the pseudospin up $%
\sigma=\Uparrow (S, 1, \frac{1}{2})$ and pseudospin down $\sigma=\Downarrow
(A, 0, \frac{1}{2})$ have the same spin. Then there is no such problem need
to be considered. Therefore, the easy-axis QHF is more likely to happen at
total filling factor $\nu=4$ than $\nu=3$.

On the basis of above discussion, we may summarize as following. The bias
gate voltage added to the sample changes the quantum well profile in the
growth direction as well as the spacial separation of the lowest two
subbands wave functions. For a larger bias gate voltage, the potential of
the well is much skewer, so the two subbands wave functions locate in the
opposite side of the quantum well (inset of Fig.~\ref{fig:d_Vg}). As a
result, the Hartree energy will arise if all the electrons stay in one
narrow subband or pseudospin level. Thus a easy-plane QHF, in which the
electrons fill the two pseudospin levels equally, is more energetically
favorable at a large bias gate voltage. In addition, a state with opposite
real spins will expend more exchange energy, so the easy-plane QHF is more
easily to form at a pseudospin configuration in which the two pseudospin
level have the same real spin (filling factor $\nu =3$ in this paper).

\section{Conclusion}

\label{sec:conclusion}

Using self-consistent local density approximation method and Hartree-Fock
mean field theory, we calculated wave function in the growth direction and
the anisotropy energy in the two-subband quantum Hall system. The data shows
great consistent with the observed easy-plane and easy-axis quantum hall
ferromagnets at filling factor $\nu =3$ and $\nu =4$ in the experiment of
Zhang \textit{et al.} \cite{Jiang2006} . Also, by analyzing the numerical
result, we give an easy-to-understand explanation about the anisotropy
occuring at these filling factors.

\section*{Acknowledgments}

\label{sec:acknowlegements}

This work was funded by National Fundamental Research Program, the
Innovation funds from Chinese Academy of Sciences, NCET-04-0587, and
National Natural Science Foundation of China (Grant No. 60121503, 10574126,
10604052).


\begin{thebibliography}{99}
\bibitem{review} R. E. Prange and S. M. Girvin (eds.), \textit{The Quantum
Hall Effect} 2nd edn (Springer, New York, 1990).

\bibitem{Eisenstein1994} S. Q. Murphy, J. P. Eisenstein, G. S.  Boebinger,
L. N. Pfeiffer, and K. W. West, Phys. Rev. Lett.  \textbf{72}, 728 (1994).

\bibitem{KunYang1994} K. Yang, K. Moon, L. Zheng, A. H. MacDonald, S.  M.
Girvin, D. Yoshioka, and S. C. Zhang, Phys. Rev. Lett.  \textbf{72}, 732
(1994).

\bibitem{Wescheider} V. Piazza, V. Pellegrini, F. Beltram, W.  Wegscheider,
T. Jungwirth, and A. H. MacDonald, Nature \textbf{402},  638 (1999).

\bibitem{Shayegan} Y. P. Shkolnikov, E. P. De Poortere, E. Tutuc, and  M.
Shayegan, Phys. Rev. Lett. \textbf{89}, 226805 (2002).

\bibitem{QHF} For a review on quantum Hall ferromagnets, see  experimental
chapter by J.P. Eisenstein and theoretical chapter by  S. M. Girvin and A.
H. MacDonald in S. Das Sarma and A. Pinczuk  (eds.), \textit{Perspectives on
Quantum Hall Effects} (Wiley, New  York, 1997).

\bibitem{MacDonald1990} A. H. MacDonald, P. M. Platzman, and G. S.
Boebinger, Phys. Rev. Lett. \textbf{65}, 775 (1990).

\bibitem{MacDonald2000} T. Jungwirth and A. H. MacDonald, Phys. Rev. B
\textbf{63}, 035305 (2000).

\bibitem{Shayegan1998} T. Jungwirth, S. P. Shukla, L. Smr\v{c}ka, M.
Shayegan, and A. H. MacDonald, Phys. Rev. Lett. \textbf{81}, 2328  (1998).

\bibitem{Hirayama} K. Muraki, T. Saku, and Y. Hirayama, Phys. Rev.  Lett.
\textbf{87}, 196801 (2001).

\bibitem{Jiang2006} X. C. Zhang, I. Martin, and H. W. Jiang, Phys. Rev.  B
\textbf{74}, 073301 (2006).

\bibitem{Abolfath} M. Abolfath, L. Belkhir, and N. Nafari, Phys. Rev. B
\textbf{55}, 10643 (1997).

\bibitem{LDA} F. Stern and S. Das Sarma, Phys. Rev. B \textbf{30}, 840
(1984).

\bibitem{ExchangeCorrelation} L. Hedin and B. I. Lundqvist, J. Phys. C
\textbf{4}, 2064 (1971).

\bibitem{Algorithm} J. M. Blatt, J. Comp. Phys. \textbf{1}, 382  (1967).
\end{thebibliography}
\end{document}